\documentclass[12pt]{iopart}

\usepackage{graphicx}

\newcommand\acis{Adv. Colloid Interface Sci. }
\newcommand\acp{Adv. Chem. Phys. }
\newcommand\cp{Chem. Phys. }
\newcommand\jcis{J. Colloid Interface Sci. }
\newcommand\jcp{J. Chem. Phys. }
\newcommand\jpa{J. Phys. A }
\newcommand\jpcm{J. Phys. Cond. Mat. }
\newcommand\jsp{J. Stat. Phys. }
\newcommand\molp{Mol. Phys. }
\newcommand\pre{Phys. Rev. E }
\newcommand\prl{Phys. Rev. Lett. }
\newcommand\sci{Science }

\begin{document}

\title[Modelling colloids with Baxter's adhesive hard sphere model]
{Modelling colloids with Baxter's adhesive hard sphere model}
\author{M A Miller\dag\ and D Frenkel\ddag}
\address{\dag\ University Chemical Laboratory, Lensfield Road, Cambridge CB2 1EW, U.K.}
\address{\ddag\ FOM Institute for Atomic and Molecular Physics, Kruislaan 407, 1098 SJ Amsterdam, The Netherlands}
\eads{\mailto{mam1000@cam.ac.uk}, \mailto{frenkel@amolf.nl}}

\begin{abstract}
The structure of the Baxter adhesive hard sphere fluid is examined using computer simulation.
The radial distribution function (which exhibits unusual discontinuities due to the particle
adhesion) and static structure factor are calculated with high accuracy over a range of conditions
and compared with the predictions of Percus--Yevick theory.  We comment on rigidity in percolating
clusters and discuss the role of the model in the context of experiments on colloidal systems
with short-range attractive forces.
\end{abstract}

\section{Introduction}

In contrast to atoms and small molecules, colloidal particles often interact through forces
whose range is much shorter than the size of the particles themselves.  The nature of such
colloids and the origins of the forces are diverse, but the qualitatively similar form of
the attraction leads to some common properties.
\par
One of the most appealing systems of this type consists of hard spherical particles suspended
in a solution of free polymer.  An effective attraction of entropic origin arises between
the spheres when they approach close enough to exclude the polymer from between them.
This depletion effect can lead to colloidal phases analogous to all three of those expected
in an atomic system: gas (low-density fluid), liquid (high-density fluid) and crystal
\cite{Ilett95a}.  The range and strength of the attraction can be varied continuously and
independently by adjusting, respectively, the concentration and size of the polymer.
The fully tunable character of the depletion
interaction provides experiments with a flexibility that is not available in atomic systems
and had previously been the privilege of theory and simulation \cite{Frenkel02a}.
Confocal microscopy further
enables particle-by-particle determination of structure, providing wide scope for detailed interactions
between experiment, theory and simulation.
\par
Although there are firm theoretical foundations for adopting system-specific functional forms for
the interaction potential between colloidal particles, such as that of Asakura and Oosawa \cite{Asakura54a}
in the case of depletion, it is tempting to simplify the problem as far as possible by retaining only the
generic features of a hard repulsive core and short-range attractive tail.  For this reason,
colloidal systems are often modelled using a narrow square well potential.  The phase diagram of the
square well system is determined by the ratio of the well width to the hard core diameter,
which is typically a few per cent in colloidal applications.
\par
In 1968, Baxter introduced a model with a hard core and short-range attraction that is even
simpler than the square well fluid in the sense that there is no range parameter \cite{Baxter68a}.
Baxter considered the square well in a limit where the well depth, $\epsilon$, becomes infinite while
the well width, $d-\sigma$, becomes infinitesimal, $\sigma$ being the hard core diameter and $d$ the
outer diameter of the well.  The
resulting potential, $U(r)$, is most easily defined by the corresponding Boltzmann factor as a function
of the particle separation, $r$:
\begin{equation}
\exp[-U(r)/kT] = \Theta(r-\sigma) + \frac{\sigma}{12\tau}\delta(r-\sigma).
\label{boltzmann} 
\end{equation}
The first term in Equation (\ref{boltzmann})
is a step function that forbids hard core overlap, while the second
is a Dirac delta function that makes binary contacts energetically favourable
to an extent determined by the parameter $\tau$.
Despite the apparently infinite strength of the attraction, the integrated weight
of bound configurations of two adhesive hard spheres remains finite, and there is
a thermodynamic equilibrium between bound and unbound states.  Adhesion dominates
at low $\tau$, while ordinary hard spheres are recovered in the limit $\tau\to\infty$.
We note that the thermal energy, $kT$, does not appear on the right-hand side of
Eq.~(\ref{boltzmann}), but that $\tau$ may be regarded as an effective temperature.
Indeed, Piazza \etal found that the phase behaviour of lysozyme, electrostatically
screened in a salt solution, is well described by the Baxter model if $\tau$
is taken to depend linearly on $T$ \cite{Piazza98a}.  A qualitatively different
relationship can arise from other types of forces, and Mallamace \etal found
$\tau\propto1/T$ for a micellar system \cite{Mallamace00a}.
In many colloidal systems, temperature is not the most relevant variable.
In the case of depletion interactions, for example, the strength of the colloidal attraction
is determined by the polymer concentration, and $\tau$ is a decreasing function of this
variable \cite{Ramakrishnan02a}.
\par
These examples do not exhaust the wide variety of systems to which the Baxter model has
been applied.  Two other contrasting applications are emulsions of water droplets in
oil \cite{Chen94a} and `hairy spheres' \cite{Verduin95a}.  In the latter case, hard silica
particles are grafted with short stearyl alcohol chains, and attraction between the spheres
is induced when they are immersed in a solvent that is unfavourable for the chains,
since aggregation excludes the solvent from between the particles.
\par
A common device for mapping experimental results onto the Baxter model is to match the
second virial coefficients.  From Eq.~(\ref{boltzmann}) ones obtains
$B_2=B_2^{\rm HS}[1-1/(4\tau)]$, where $B_2^{\rm HS}=2\pi\sigma/3$ is the hard sphere
second virial coefficient.  Having obtained $\tau$, it is normal to interpret the
experimental data by reference to the results of Percus--Yevick theory for the model.
The phase diagram can be obtained through various routes
of Percus--Yevick theory, the most commonly cited being the compressibility equation
\cite{Baxter68a,Barboy74a}, which predicts that the model undergoes a first-order phase
transition between two fluid phases of different density.  The energy equation also
predicts such a transition \cite{Watts71a,Barboy79a}, but the positions of the coexistence
curve and the critical point are very different from the compressibility results.
\par
A lack of definitive knowledge of the model's phase behaviour would clearly restrict
its utility in the interpretation of experimental data.  In recent work, we have used
computer simulations to find numerical values for the critical point \cite{Miller03a},
coexistence curve, and equation-of-state \cite{longsticky} of adhesive hard spheres.
We will refer to this work in the present contribution, but focus here on the structure
of the fluid, providing high quality radial distribution functions and the corresponding
static structure factors under a variety of conditions.  The singular potential leads
to unusual features in these functions.

\section{Simulations}

It is difficult to perform equilibrium simulations of systems with short range
forces, since the energetically accessible fraction of configuration space decreases
rapidly with narrowing of the attractive well.  Exploring the relevant subspace
ergodically by conventional molecular dynamics or Monte Carlo methods becomes a
technical challenge.
\par
The infinitesimal range of the Baxter potential has allowed a specialized approach
to be devised for this model.  Two particles only attract when they are precisely in
contact, and the contact defines a spherical surface of radius $\sigma$ if one of
the particles is considered fixed and the remaining two degrees of freedom are
explored.  The total Boltzmann weight of the bound configuration
can be found by integrating Eq.~(\ref{boltzmann}) over this surface.  Because the
spherical surface has zero thickness, it has no overlap with similar surfaces that
describe contacts with other particles.  Pursuing this property, it is possible
to evaluate the weights of different bound states independently and then compare them.
In contrast, the contact zone between two square-well particles is a spherical shell
of finite thickness, and the overlap of such shells carves out increasingly complicated
shapes as more shells become involved, making it difficult to find their volumes.
\par
The possibility of comparing integrated Boltzmann weights has been exploited in a
Monte Carlo algorithm dedicated to the adhesive hard sphere model, that moves
particles by explicitly making and breaking contacts rather than by conventional random
displacements \cite{Seaton86a,Seaton87a,Kranendonk88a}.  Our own implementation
is described in detail in Ref.~\cite{longsticky}.  Although the specialized
algorithm is considerably more complicated than the standard Monte Carlo method,
it can be used to sample ergodically in the limit of infinitesimal attractive
range more efficiently than is possible using standard techniques on a
potential of very narrow, but finite, range.
\par
\begin{figure}
\centerline{\includegraphics[width=150mm]{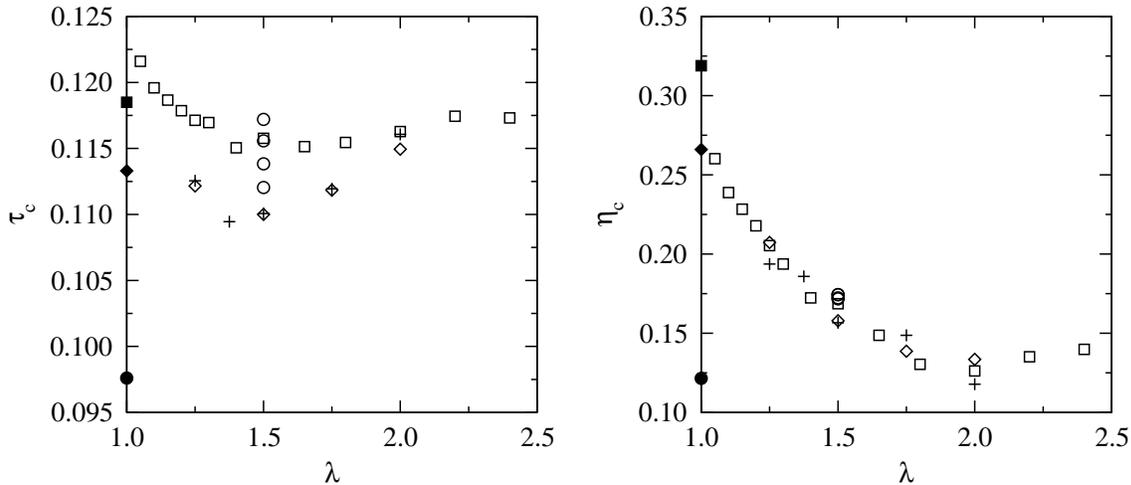}}
\caption{Critical effective temperature $\tau_c$ (left) and volume fraction $\eta_c$ (right)
of square-well fluids with range parameter $\lambda$.  Open symbols are taken from References
\cite{Brilliantov98a} (circles), \cite{Lomakin96a} (squares), \cite{Rio02a} (diamonds) and
\cite{Vega92a} (plusses).  Filled symbols at $\lambda=1$ are the adhesive hard sphere limits
from the Percus--Yevick compressibility (circles) and energy (squares) routes, and from
simulation (diamonds).
\label{limits}}
\end{figure}

It is expected, however, that since the adhesive hard sphere model was originally
derived from a square-well potential, the properties of the square-well fluid should
approach those of Baxter's model as the range parameter, $\lambda=d/\sigma$,
tends to unity.  In Figure \ref{limits} we investigate this expectation for the
case of the fluid--fluid critical point.  The critical point in the Baxter limit was obtained
numerically in previous work \cite{Miller03a}, while the square well data are taken from a number
of simulation studies in the literature \cite{Brilliantov98a,Lomakin96a,Rio02a,Vega92a}.
In the right-hand panel of Figure \ref{limits}, the square-well critical volume fractions
extrapolate smoothly to the Baxter limit of $0.265\pm0.005$.  Note that this value lies between
the predictions of the compressibility and energy routes of Percus--Yevick theory \cite{Watts71a},
also marked in the Figure, and that these predictions differ by well over a factor of two.

To compare the critical temperatures, we need a method for mapping Baxter's $\tau$ parameter
onto the thermodynamic temperature $T$ of the square well.  The most natural way to do this
is probably through Baxter's expression $\epsilon/kT=-\ln[12\tau(1-\sigma/d)]$ that couples the
square-well width and depth to define the adhesive limit \cite{Baxter68a}.
Writing $\lambda=d/\sigma$ and rearranging, we obtain
\begin{equation}
\tau^{-1} = 12(1-\lambda^{-1})e^{\epsilon/kT}.
\label{mapping}
\end{equation}
The more general approach of equating the second virial coefficients yields a different
expression that has the same form in the limit $\lambda\to1$.  The left-hand panel of
Figure \ref{limits} shows that, under the mapping of Equation (\ref{mapping}), there is
remarkably little variation in the effective critical temperature $\tau_c$ with square-well
range $\lambda$.
Extrapolating the data of Lomakin \etal \cite{Lomakin96a} to $\lambda=1$ in the left-hand
panel of Figure \ref{limits} leads to a $\tau_c$ above the predictions of both routes of
Percus--Yevick theory, as noted in Reference \cite{Lomakin96a}.  The more recent
data of del R{\'\i}o \etal lie systematically lower,
and our simulation value \cite{Miller03a} for the Baxter limit of $\tau_c=0.1133\pm0.0005$ is
somewhat below the Percus-Yevick energy
result.  We have observed that the simulated coexistence curve of the Baxter model suffers
from strong finite-size effects near the critical point \cite{Miller03a}, and therefore
employed a careful scaling analysis \cite{Bruce92a,Wilding92a,Wilding95a} to find
the infinite-system $\tau_c$.  The apparent critical temperature of a finite system is
generally an overestimate of the thermodynamic limit, and if finite-size effects increase
as the Baxter model is approached, they might account for the observed overshoot
in some of the square-well data as $\lambda\to1$.  Backing off from the adhesive sphere
limit, however, it is clear that the Percus--Yevick compressibility result of
$\tau_c=0.0976$---the value that is most often assumed---is a significant underestimate.

\section{Structure of the adhesive hard sphere fluid\label{structure}}

\begin{figure}
\centerline{\includegraphics[width=150mm]{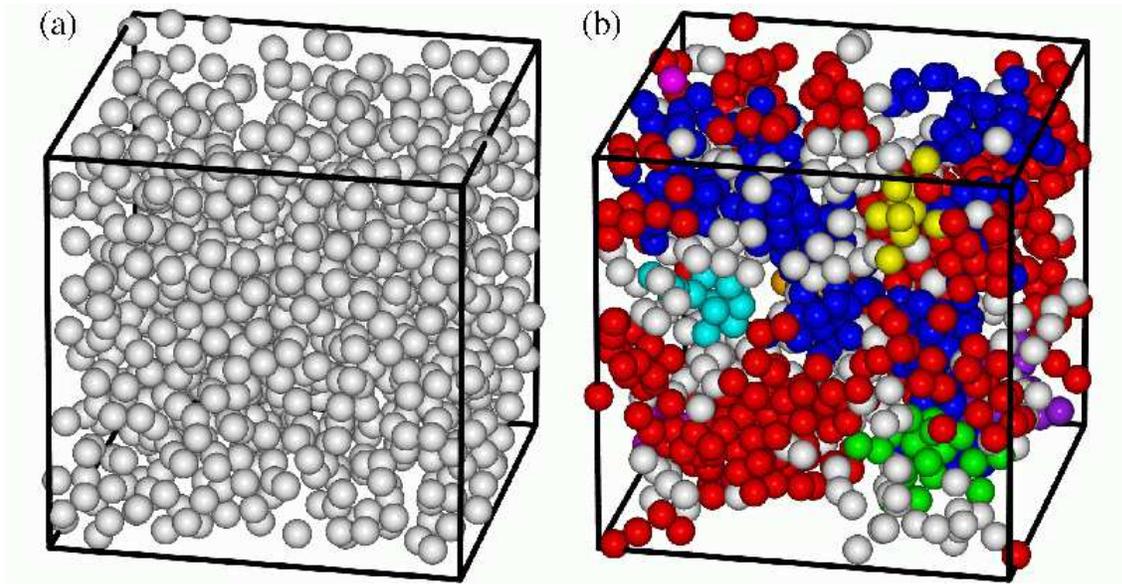}}
\caption{Simulation snapshots of $N=864$ particles at a volume fraction of $\eta=0.164$.  Left:
ordinary hard spheres, right: adhesive hard spheres at $\tau=0.13$.  The largest clusters
have been highlighted in colour.  The red cluster percolates.
\label{snapshots}}
\end{figure}

Figure \ref{snapshots} illustrates some instantaneous configurations from simulations
of $N=864$ particles at a fairly low volume fraction of $\eta=0.164$.  (The volume
fraction is related to the number density $\rho=N/V$ by $\rho\sigma^3=6\eta/\pi$.)  In
the hard-sphere case, the particles fill the container uniformly.  In the snapshot of
adhesive hard spheres at $\tau=0.13$, however, particles have aggregated into a number
of clusters that are locally
dense, leaving sizeable voids in the overall structure.  A cluster is unambiguously
defined as a set of particles connected by an unbroken chain of contacts.  The largest
cluster in the snapshot spans the simulation cell in the sense that it connects periodic
images of the same particle once the periodic boundary conditions have been taken into
account.  Such a cluster is said to percolate and is the analogue of an infinite cluster
in an experimental system.  In some systems, the onset of percolation can be detected by
a sudden change in a physical property, such as the electrical conductivity in a
microemulsion \cite{Chen94a}.  If one defines a gel to be a non-compact space-filling
structure of particles \cite{Poon97a}, then a system of adhesive hard spheres begins to exhibit
gel-like properties when the volume fraction crosses the ($\tau$-dependent) percolation threshold
and infinite clusters are always present.
However, it is important to remember that all clusters are dynamic, since particles are not
irreversibly bound.  Indeed, the conditions of Figure \ref{snapshots}(b) lie on the percolation
threshold itself \cite{longsticky}, where the lifetime of a percolating cluster is very short and
may be determined by the removal of a single particle.

\begin{figure}
\centerline{\includegraphics[width=100mm]{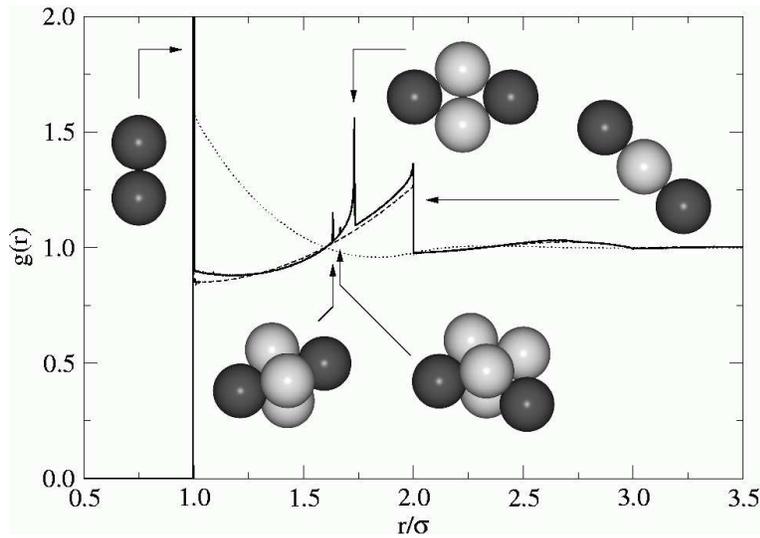}}
\caption{
Radial distribution function, $g(r)$, at a volume fraction of $\eta=0.164$
as for Figure \ref{snapshots}.
Solid line: adhesive hard spheres at $\tau=0.13$ from simulation; dashed line:
the corresponding Percus--Yevick prediction; dotted line: hard spheres.
Physical clusters contributing to certain features of the solid line are
indicated.
\label{demo}}
\end{figure}

The effect of surface adhesion on the structure of the fluid is more clearly seen in the
radial distribution function, $g(r)$, which is shown in Figure \ref{demo} for the same
conditions as the snapshots in Figure \ref{snapshots}.  At this low volume fraction, the hard
sphere fluid (dotted line) shows little structure.  In contrast, the adhesive hard sphere
fluid has a number of striking features.  Physical clusters in which the separation of
two particles cannot be changed without breaking a contact contribute a delta-function
to $g(r)$, since a single geometry then receives a finite statistical weight.
The delta-functions appear to have finite height in Figure \ref{demo} due to the finite
bin width ($0.002\sigma$) used in accumulating $g(r)$.  The smallest
clusters that make delta-function contributions at distinct distances contain 2, 5 and 6
particles, as illustrated in Figure \ref{demo}.  The darker particles indicate the separations
corresponding to the peaks, which are located at $r/\sigma=2$, $\sqrt{8/3}$ and $5/3$.
The tendency for particles to bind, which gives rise to the strong delta-function at
$r=\sigma$, also explains the contrasting trend in $g(r)$ from ordinary hard spheres
in the range $1<r/\sigma<2$: the presence of a tightly-bound shell of nearest neighbours
excludes other particles from this range, resulting in a deficit of particles close to
$r=\sigma$ relative to the bulk density.  The surplus exhibited by hard spheres indicates
a more loosely structured coordination shell arising purely from excluded volume
considerations.

In addition to the singularities, $g(r)$ has two discontinuities in the range plotted
in Figure \ref{demo}.  Discontinuities arise when the distance between two particles
in a flexible cluster is extremized with respect to variation of the cluster geometry
while the topology of contacts is held fixed.  Consider, for example, the trimer
illustrated in Figure \ref{demo}.  Without breaking or making any
contacts (and therefore without changing the energy), the cluster can be continuously
deformed between an equilateral triangle, where
the dark particles are separated by $\sigma$, to the linear structure shown, where the
separation is $2\sigma$.  A greater separation would require the breaking of a bond.
Similarly, the tetramer can be deformed continuously to give separations between $\sigma$
and $\sqrt{3}$ without altering the topology of contacts.
\par
Cummings \etal have noted that, within the framework of Percus--Yevick theory for this
model, a discontinuity in the
$n^{\rm th}$ derivative of the radial distribution function at $r$ gives rise to a
discontinuity in the $(n+1)^{\rm th}$ derivative at $r+\sigma$ \cite{Cummings76a}.
Hence, the delta-function at $r=\sigma$ gives rise to the discontinuity at $r=2\sigma$,
which in turn causes the discontinuous gradient at $r=3\sigma$, also reproduced
by the simulations.  Percus--Yevick theory, however, neglects certain cluster
diagrams, causing it to miss many of the other delta-functions and discontinuities
accounted for in the simulations.  Glandt and coworkers have provided expressions for many
low-order cluster integrals, and used them to calculate cluster concentrations and
delta-function coefficients for $g(r)$ \cite{Post86a,Seaton87a}.
\par
\begin{figure}
\centerline{\includegraphics[width=150mm]{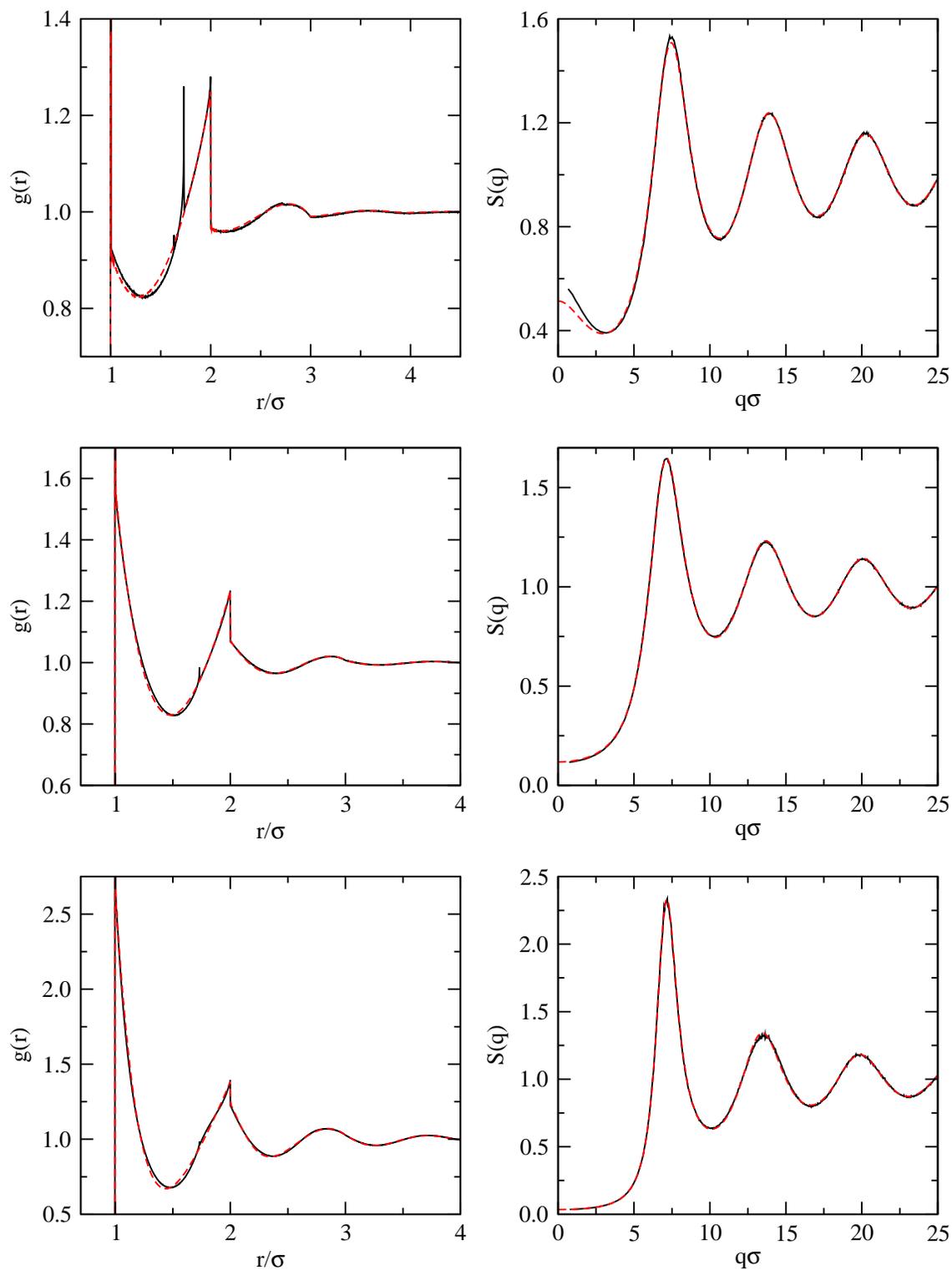}}
\caption{
Radial distribution function (left) and static structure factor (right) under
three sets of conditions: $\tau=0.2$, $\eta=0.32$ (top row); $\tau=0.5$, $\eta=0.4$
(middle row); $\tau=1$, $\eta=0.5$ (bottom row).  Solid black lines are from simulation
and dashed red lines from Percus--Yevick theory.
\label{survey}}
\end{figure}

We are now in a position to calculate structural properties of the adhesive hard
sphere fluid more precisely than was previously possible \cite{Kranendonk88a}.  In
Figure \ref{survey} we present the radial distribution function and the static
structure factor, $S(q)$, for a series of conditions considered by Kranendonk and
Frenkel \cite{Kranendonk88a}, although we omit the combination $\tau=0.1$, $\eta=0.14$
as this lies within the spinodal region of the phase diagram.  $S(q)$ is related to
$g(r)$ by a Fourier transform, but is calculated in the simulations directly from
$S(q)=\langle\rho({\bf q})\rho(-{\bf q})\rangle$, where
\begin{equation}
\rho({\bf q})=\sum_i^N \exp({-{\bf q}.{\bf r}_i})
\end{equation}
and ${\bf r}_i$ is the position of particle $i$.  Because of the periodic boundary
conditions, it is only possible to study fluctuations with
${\bf q}=2\pi L^{-1}(q_x, q_y, q_z)$  where $q_x$, $q_y$, and $q_z$ are all integers
and $L$ is the length of the cubic simulation cell.  2000 ${\bf q}$-vectors with
randomly-chosen orientation were sampled, covering the range plotted approximately
uniformly.

Notwithstanding the failure of Percus--Yevick theory to capture all but one of
the delta-functions
labelled in Figure \ref{demo}, the agreement between theory and simulation is good,
and improves as $\tau$ increases, corresponding to less dominant surface adhesion.
The delta-functions, however, are a defining feature of the adhesive hard sphere model,
and the omission of the corresponding cluster integrals is presumably one of the reasons
for the significant difference between the equations of state derived through different
functionals from the Percus--Yevick $g(r)$.
\par
The theory does remarkably well for the prefactor of the delta-function at
$g(\sigma)$, which is directly proportional to the average coordination number
(number of contacts per particle).  The Percus--Yevick coordination number is available
in simple closed form \cite{Chiew83a} and under the three sets of conditions displayed
in Figure \ref{survey} takes the values 2.93, 2.51 and
2.65 respectively, to be compared with 2.97, 2.48 and 2.69 from simulation (the latter
being converged to within the precision quoted).  For a very recent analysis of
adhesive hard spheres in the theoretical framework of a generalized closure that
subsumes the Percus--Yevick see Reference \cite{Gazzillo04a}, which includes comparisons
with the data in Figure \ref{survey}.

\section{Rigidity}

The percolation threshold denotes the set of conditions on the phase diagram at
which the cluster size
diverges, or in a simulation, at which system-spanning clusters are observed with a
probability of 50\%.  (The transition to percolation is sharp, and the 50\% criterion,
though arbitrary, is robust.)  As the threshold is crossed, thermodynamic properties
and their derivatives change smoothly, and so percolation is not a thermodynamic
phase transition.  The connectivity of the particles may, however, lead to sudden
changes in other properties such as electrical conductivity \cite{Chen94a}.
\par
A percolating cluster is defined only by its connectivity through the system.  In
principle, a connected cluster with only a minimum of connections can be deformed
without making or breaking any contacts between particles and therefore without
any energetic penalty.  In contrast to this `bond percolation' one might expect a
change in mechanical properties if a percolating cluster contained a rigid backbone
that, due to its network of contacts, cannot be deformed without altering its topology.
To find whether a cluster exhibits such `rigidity percolation' one must discount
any `floppy' (non-rigid) modes, and determine whether the structure remains
constrained \cite{Jacobs95a}.  Because of the additional contacts required to
constrain the geometry, any rigidity percolation threshold is expected to lie to the
high-density side of the bond percolation threshold on the phase diagram.
\par
In a system of adhesive hard spheres, the smallest rigid subunit is a maximally
connected tetrahedron, which possesses six contacts.  Larger rigid structures can
be produced by adding triply-coordinated particles to the faces of the tetrahedron,
producing a network of face-sharing tetrahedra.  Two tetrahedra may be connected
by a common edge or vertex, but such connections lend flexibility to the overall cluster.
We may therefore determine the extent of rigidity by searching for face-sharing polytetrahedral
networks.
\par
\begin{figure}
\centerline{\includegraphics[width=110mm]{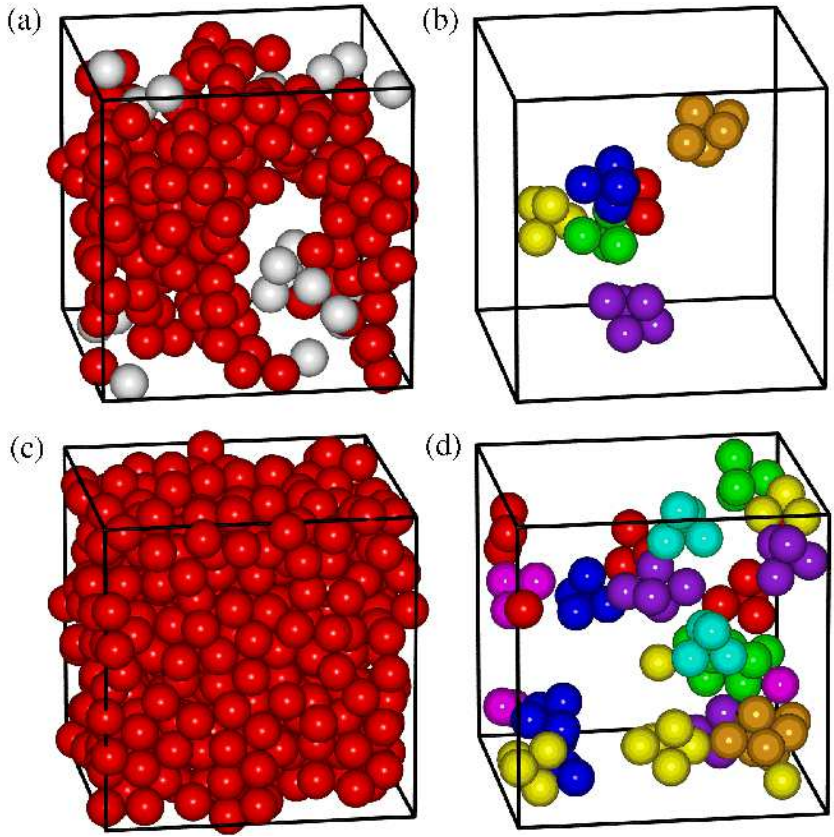}}
\caption{
Snapshots from grand canonical simulations \cite{longsticky} at $\tau=0.13$.
All particles are shown in
the left-hand frames, with particles in the percolating cluster coloured red.  On the
right, any particles not belonging to a rigid subunit have been removed, and different
colours are used to distinguish between the subunits.  Frames (a) and (b) are taken
from simulations with an average packing fraction of 0.17, and (c) and (d) an
average of 0.50.
\label{rigidity}}
\end{figure}

Figure \ref{rigidity} illustrates some snapshots from a rigidity analysis of
adhesive hard spheres at $\tau=0.13$ and two volume fractions, both of which
lie within the bond-percolated regime.  As the left-hand panels show, most particles
belong to the bond-percolating cluster.  In the right-hand panels, the rigid
subunits are displayed in different colours, and particles possessing independent
unconstrained degrees of freedom are not shown.  It is clear that, even at a
high volume fraction of $\eta=0.50$, rigidity does not extend further than a few
particles.  Several rigid tetrahedra share one or two particles (in which case
the assignment of the shared particles to either tetrahedron is arbitrary), but
it is rare to see a fully constrained unit of more than six particles.

As the volume fraction of the system approaches its maximum, the physical space
available to floppy modes tends to zero, and mechanical properties should change
accordingly.  However, as we find no evidence for a rigidity percolation threshold
in this system, we expect these changes to occur smoothly rather than showing any
sharp transitions.

\section{Some remarks: gels and glasses}

Finally, we comment on the adhesive sphere fluid in the context of
recent work on glasses.  Mode-coupling theory for particles with sufficiently
short-range attractive forces predicts
two types of structural arrest: a `repulsive' glass in which particles are
trapped within the local cage of their neighbours at high volume fraction, and an
`attractive glass' where particles are energetically bound to their neighbours
even at lower volume fraction.  There is a re-entrant glass--liquid--glass transition
as a function of adhesive strength at constant volume fraction and a line of
glass--glass transitions extending from the point where the two liquid--glass
lines meet \cite{Dawson01a}.  These predictions are borne out in
experiments \cite{Pham02a} and simulations \cite{Puertas02a}.
\par
Being the archetypal short-range attractive system, it seems natural to apply the
mode-coupling analysis to the Baxter model.  There are, however, at least two
obstacles to doing this directly.  From the simulation approach, Monte Carlo
techniques allow us to probe the thermodynamics and equilibrium structure of
adhesive hard spheres, but do not provide rigorous information on dynamic processes.
Since the time-scales for bond making and breaking diverge as the adhesive limit
is approached \cite{Stell91a}, direct molecular dynamics simulations are not
feasible.  From the point of view of mode-coupling theory, the Baxter limit is
problematic because the structure factor, which is the main input to the theory,
contains a slowly decaying ($1/q$) tail that makes the integral over $q$ in the
mode-coupling functional
diverge.  It is therefore necessary to introduce an arbitrary reciprocal-space
cutoff $q_{\rm max}$ upon which the results depend \cite{Dawson01a,Fabbian99a}.
\par
\begin{figure}
\centerline{\includegraphics[width=90mm]{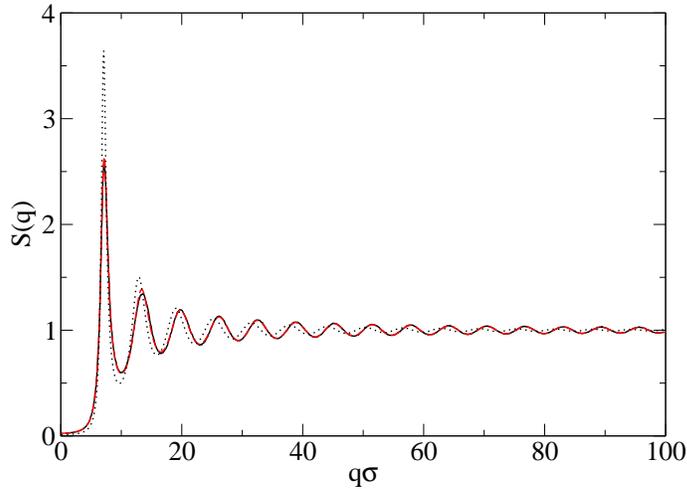}}
\caption{
Static structure factor of adhesive hard spheres at $\tau=1.32$ and $\eta=0.52$,
illustrating its slowly-decaying $1/q$ tail.
The solid black line is from simulation and the dashed red line from Percus--Yevick
theory.  For comparison, the dotted line shows the simulation result for ordinary
hard spheres at the same volume fraction.
\label{longsofk}
}
\end{figure}

Considering the static structure factor as the Fourier transform of the radial
distribution function, the $1/q$ term in $S(q)$ arises specifically from the
delta-function at $r=\sigma$ in $g(r)$.  The slowly-decaying tail is well
reproduced by simulation, as illustrated in Figure \ref{longsofk} for $\tau=1.32$,
$\eta=0.52$.  These conditions are close to the region of the phase diagram
where the attractive and repulsive glass lines meet \cite{Dawson01a}.  The
plot also serves to demonstrate that Percus--Yevick theory works well for $S(q)$
in the regime of greatest relevance to mode-coupling theory.

\begin{figure}
\centerline{\includegraphics[width=90mm]{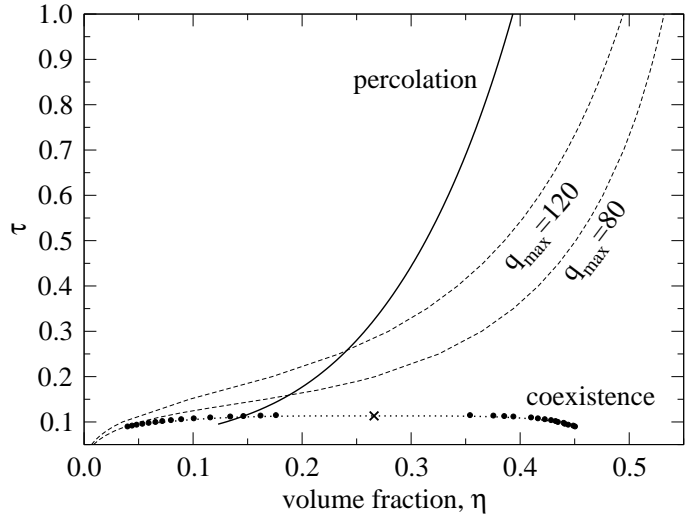}}
\caption{
Phase diagram of adhesive spheres in the $\eta$--$\tau$ plane.  The dotted line with
circles is the fluid--fluid coexistence curve from simulation, with the critical
point marked by a cross \cite{Miller03a}.  The solid line is the percolation threshold
from simulation \cite{longsticky} to the high-$\eta$ side of which infinite clusters
exist in the system.  The dashed lines are the mode-coupling attractive
glass transition with cutoffs $\sigma q_{\rm max}=80$ and 120 as marked.  Mode-coupling
theory predicts structural arrest to the high-$\eta$ side of this threshold.
\label{mct}}
\end{figure}

Figure \ref{mct} shows the phase diagram of adhesive hard spheres with the
fluid--fluid coexistence line and (bond) percolation threshold as determined
by simulation \cite{longsticky}.  Infinite clusters dominate the system to the
high-$\eta$ side of the percolation threshold.  The threshold demarks the well-defined
set of conditions where the structure of the fluid first becomes connected
on a macroscopic scale and (due to its low overall volume fraction) locally
inhomogeneous in density, unlike a normal liquid.  On structural grounds, therefore,
we may think of the percolated part of the phase diagram as a gel, even though
crossing the percolation threshold itself is not accompanied by immediate
structural arrest.  Progressing further into the percolated regime, the infinite
clusters become bolstered with more neighbours.  It then requires many bonds to
be broken simulataneously for a given cluster to stop percolating, and the
connectedness of the cluster becomes far less transitory.

For comparison, Figure \ref{mct} also shows two lines
corresponding to the attractive glass transition from mode-coupling theory
\cite{Dawson01a} with two values of the cutoff $q_{\rm max}$.  The range of
the $\tau$ axis extends to somewhat below the meeting point of the attractive
and repulsive glass transitions, and the latter therefore does not appear on this
plot.  To the high-$\eta$ side of the attractive glass line, mode-coupling
theory predicts the fluid to by structurally arrested due to the dominance of
adhesion between particles.  Although the lines are shown all the way down to low
volume fraction, mode-coupling theory is expected to be most reliable at high
volume fraction, where
the equilibrium structure factor is likely to resemble that of the arrested
phase quite closely.  In contrast, it is questionable whether the equilibrium
$S(q)$ can be used to predict structural arrest---by definition a non-equilibrium
property---at densities low enough for the structure to differ significantly from
equilibrium.
Indeed, at very low volume fraction, the average number of bonds per particle is
not sufficient to support large clusters at all \cite{Zaccarelli01a}.
\par
The two dashed curves in Figure \ref{mct} illustrate the effect of increasing
$q_{\rm max}$ in mode-coupling theory, which is to assign more of the phase diagram
to the attractive glass state.  Towards the high volume-fraction ends of these
lines (where they are expected to be most reliable), they approximately parallel
the percolation threshold.  The physical interpretation of introducing a $q$-space
cutoff is to back off from the limit of infinitesimal attractive range.  The
square-well
ranges corresponding to $\sigma q_{\rm max}=80$ and 120 may be estimated to be about
$\lambda=1.04$ and 1.026, respectively \cite{Gotze03a}; both are relevant
for experimental applications in colloids.  While the attractive-glass line should in
principle continue to advance indefinitely to lower densities as $q_{\rm max}$ is increased,
it probably makes little sense to consider anything to the low-$\eta$ side of the
percolation threshold as glassy, since in that regime the system is a fluid of
separated and fluctional clusters.  Therefore, while the percolation threshold does
not contain dynamical information, it nevertheless retains
an important utility as an unambiguous transition to a system-wide connected
state which, as we have argued, possesses at least some properties of a gel.
\par
It is also apparent from Figure \ref{mct} that the critical point for the
fluid--fluid separation of adhesive hard spheres lies both within the
percolated part of the phase diagram and inside the attractive glass region
even for $q$-space cutoffs somewhat smaller than $80/\sigma$ (and corresponding ranges
longer than $\lambda=1.04$).  The assertion made in previous work \cite{Miller03a},
that gelation is likely to interfere with equilibrium phase separation in systems
with short-range attractive forces, therefore remains highly plausible.

\ack
MAM is very grateful to Dr.~Matthias Sperl for the MCT data plotted in Figure
\ref{mct} and for discussions regarding the application of MCT to the
Baxter model.  The work of the FOM Institute is part of the research program of
FOM and is made possible by financial support from the Netherlands
Organization for Scientific Research (NWO).

\end{document}